\documentstyle[pre,aps,preprint,12pt]{revtex}
\begin{document}
\draft
\tightenlines

\title{Simple Model with Facilitated Dynamics for Granular Compaction}
\author{J.\ Javier Brey, A.\ Prados}
\address{F\'{\i}sica Te\'orica, Facultad de F\'{\i}sica, Universidad
         de Sevilla, Apdo.\ de Correos 1065, E-41080 Sevilla, Spain}

\author{B.\ S\'anchez-Rey}
\address{F\'{\i}sica Aplicada, E.\ U.\  Polit\'ecnica, Universidad de Sevilla,
        Virgen de \'Africa 7, E-41011  Sevilla, Spain}
\date{today}
\maketitle
\begin{abstract}
A simple lattice model is used to study compaction in granular media. As in
real experiments, we consider a series of taps separated by large enough
waiting times. The relaxation of the density exhibits the characteristic
inverse logarithmic law. Moreover, we have been able to identify analytically
the relevant time scale, leading to a relaxation law independent of the
specific values of the parameters. Also, an expression  for the asymptotic
density reached in the compaction process has been derived. The theoretical
predictions agree fairly well with the results from the Monte Carlo
simulation.

\end{abstract}

\section{introduction}
\label{sec1}
One of the characteristic complex behaviors exhibited by granular materials is
compaction  \cite{ByM92,KFLJyN95,NKBJyN98,HYRCyK94}. It can be roughly defined
as the density relaxation of a loosely packed system of many grains under
mechanical tapping or vibration. Granular compaction is important to many
industrial applications related with the production and manipulation of a wide
variety of systems composed by many macroscopic particles or grains
\cite{KFLJyN95}. In the last few years, a series of experiments have been
carried out trying to identify the physical principles underlying granular
compaction \cite{KFLJyN95,NKBJyN98,NKPJyN97}. Starting from a loosely packed
initial configuration, systems of monodisperse glass beads were tapped
vertically. The waiting time between successive taps was large enough to allow
the system to relax, so that the beads were at rest before the next tap started.
The time evolution of the density towards a steady state has been analyzed, and
it has been shown that it can be accurately described by an inverse logarithmic
law with four adjustable parameters, whose values depend only on the tapping
strength measured  by the peak acceleration of a tap. The logarithmic relaxation
has been found in many different models \cite{CLHyN97,NCyH97,BKNJyN98}
suggesting that such a behavior is quite general \cite{Ja98}. Although several
mechanisms have been proposed to explain the behavior observed in the
experiments, a fully satisfactory theory is still lacking.

Here we consider a one-dimensional model  simple enough as to allow some
detailed calculations. One of our aims was to try to identify the relevant time
scale over which the relaxation (compaction) of the system takes place. This is
the first step in the search of general laws governing the physics of
densification. A main difficulty in studying compaction is that there are two
different series of elementary processes involved in the experiment. The system
is submitted to taps or pulses separated by time intervals for which the system
is allowed to relax freely. The initial state for each tap is the final state
from the previous relaxation. Both processes, tapping and free evolution, must
be considered in detail, and they are rather different from a physical point of
view. For instance, while the duration of the pulse is clearly a relevant
parameter of the problem, the free relaxation is assumed to last by definition
until the system gets trapped and it is at rest. Quite interestingly, the
experiments have shown that it is useful to measure time by the number of pulses
applied to the system.  Another central question is how much settling will occur
for a given vibration intensity, and also if the stationary value of the density
depends on the initial configuration. How these facts appear in our model and
which is the role played by the duration and amplitude of the pulses are points
we will address here.

The plan of the paper is as follows. In the next Section the model will be
presented. It consists of a lattice whose sites can be occupied by particles.
The dynamics is formulated by  means of a master equation and it is facilitated,
in the sense that the rates of adsorption and desorption of a particle are
proportional to the number of particles in the nearest neighbor sites. The model
can be exactly solved in the no desorption limit, which corresponds to the very
low temperature limit. The solution is obtained in Sec.\ \ref{sec3} and
describes the evolution of the system without external perturbation. Therefore,
it will be used to study the relaxation of the system towards a metastable state
between pulses. In spite of the simplicity of the system, the general solution
for arbitrary strength of the external energy source is rather complicated. We
have considered the limit of short duration of the taps, not only because of
mathematical convenience, but also because it seems to be the limit in which the
time scales involved in the problem become well separated.

The sequence of taps and free relaxation processes, i.\ e.\ compaction is the
subject of Sec.\ \ref{sec4}. An expression for the density after the $n+1$ tap
in terms of the density and the probability distribution of two holes separated
by a site after the previous tap is derived. Although this relation does not
provide an explicit expression for the evolution of the density, it allows to
identify the relevant time scale, which turns out to be proportional to the
duration of a tap times a parameter measuring their strength. Curves describing
the density evolution of systems starting from the same initial state but
corresponding to different values of the parameters are shown to be the same
when plotted as functions of the scaled time. Besides, the single scaled curve
is very well fitted by the inverse logarithmic law known from real experiments.

In the limit of many taps, the density reaches a steady value that is discussed
in Sec.\ \ref{sec5}.  By using a pair approximation it is found that the steady
density is proportional to the time relevant parameter mentioned above. This
prediction agrees well with the numerical results from the simulations. To the
best of our knowledge this is the first time an explicit expression for the
asymptotic density limit in a discrete tapping process has been analytically
derived from a model for compaction. Finally, the last section contains some
final remarks and comments.

\section{Description of the model}
\label{sec2}
We consider a one-dimensional lattice with $N$ sites. Each site can be either
occupied by a particle or empty. A configuration of the system is specified, for
instance, by giving an ordered sequence of N particles and holes. Let us
introduce a set of variables ${\bf m} \equiv \{ m_{i}; i=1,2,\cdots,N\}$, such
that $m_{i}$ vanishes if there is a particle at site $i$, while it takes the
value $1$ if there is a hole, i.\ e.\ the site $i$ is empty.

The dynamics of the system is defined as a Markov process and formulated
by means of the master equation for the conditional probability
$p_{1/1}({\bf m},t|{\bf m}^{\prime},t^{\prime})$ of finding the system in
the configuration ${\bf m}$ at time $t$, given it was in the configuration
${\bf m}^{\prime}$ at time $t^{\prime}<t$ \cite{vk92},
\begin{equation}
\label{2.1}
\frac{\partial}{\partial t}\, p_{1/1}({\bf m},t|{\bf m}^{\prime},t^{\prime})
=\sum_{i}\left[ W_{i}(R_{i}{\bf m})p_{1/1}(R_{i}{\bf m},t|{\bf m}^{\prime},
t^{\prime})-W_{i}({\bf m})p_{1/1}({\bf m},t|{\bf m}^{\prime},t^{\prime})
\right],
\end{equation}
where $R_{i}{\bf m} \equiv \{m_{1},\cdots,R_{i}m_{i},\cdots,m_{N}\}$
with $R_{i}m_{i}=1-m_{i}$, i.\ e.\,  $R_{i}{\bf m}$ is the configuration
obtained from ${\bf m}$ by changing the state of hole or particle of site
$i$. The above equation is to be solved with the initial condition
\begin{equation}
\label{2.2}
p_{1/1}({\bf m},t^{\prime}|{\bf m}^{\prime},t^{\prime})=
\delta_{{\bf m},{\bf m}^{\prime}}=\prod_{i=1}^{N}\delta_{m_{i},
m^{\prime}_{i}}.
\end{equation}
The one--time distribution
\begin{equation}
\label{2.2a}
p({\bf m},t)=\sum_{{\bf m}^{\prime}} p_{1/1}({\bf m},t|{\bf m}^{\prime},0)
p({\bf m}^{\prime},0)
\end{equation}
also obeys Eq.\  (\ref{2.1}), although now the initial condition must be
given in each specific situation.

The possible elementary processes occurring in the system are the adsorption of
a particle on an empty site from a surrounding bulk and the desorption of a
particle from the lattice to the bulk. The probability that an adsorption
attempt be made on site $i$ in the infinitesimal time interval between $t$ and
$t+d t$ is $k_{+}dt$. Of course, a particle can  be adsorbed only if the site is
empty. In the same way, the probability per unit of time that a given particle
try to leave the lattice is $k_{-}$. Both processes are restricted in the
following way. A particle can be adsorbed on or desorbed from a site only if at
least one of its nearest neighbor sites is empty. More precisely, the
probability rate for the events is proportional to the number of nearest
neighbor holes. This condition tries to model naively the short-ranged
geometrical constraints that make structural rearrangements difficult in a
granular material. Therefore, we assume that the transition rates are given by
\begin{equation}
\label{2.3}
W_{i}({\bf m}^{(i)},m_{i}=1)=k_{+}\frac{m_{i-1}+m_{i+1}}{2}\,,  \quad
W_{i}({\bf m}^{(i)},m_{i}=0)=k_{-}\frac{m_{i-1}+m_{i+1}}{2}\, ,
\end{equation}
with ${\bf m}^{(i)}=\{m_{1},\cdots,m_{i-1},m_{i+1},\cdots, m_{N}\}$. A similar
kind of facilitated dynamics has been used previously in the context of Ising
models \cite{FyA84}. Although we restrict ourselves here to the one-dimensional
case, the model can be formulated for arbitrary dimension. Let us introduce new
constant parameters $\nu$ and $\epsilon$ by
\begin{equation}
\label{2.4}
\nu= k_{-}+k_{+}, \quad  \epsilon=\frac{k_{-}}{k_{-}+k_{+}}.
\end{equation}
The constant $\nu$ has the dimensions of a frequency and $\epsilon$ is a
dimensionless parameter defined in the interval $0\leq \epsilon \leq 1$. For
$\epsilon=1$ no particle is adsorbed by the system, while for $\epsilon=0$
desorption processes do not occur. In terms of these parameters, Eqs.\
(\ref{2.3}) can be written together as
\begin{equation}
\label{2.5}
W_{i}({\bf m})=\frac{\nu}{2}(m_{i-1}+m_{i+1})[\epsilon+m_{i}(1-2\epsilon)].
\end{equation}
The ratio of the desorption transition rates to the adsorption ones is
\begin{equation}
\label{2.6}
\frac{W_{i}({\bf m}^{(i)},m_{i}=0)}{W_{i}({\bf m}^{(i)},m_{i}=1)}=
\frac{k_{-}}{k_{+}}=\frac{\epsilon}{1-\epsilon}=x,
\end{equation}
where the last equality defines the parameter $x$.

The stochastic process we have formulated has a steady one-time distribution
of the form
\begin{equation}
\label{2.7}
p_{st}({\bf m})=\frac{1}{(1+x)^{N}}\prod_{i=1}^{N}x^{m_{i}}\ ,
\end{equation}
and the density of holes (average number of holes divided by the total number of
sites $N$) in the steady state is
\begin{equation}
\label{2.8}
\langle m_{i}\rangle_{st}=\sum_{\bf m}m_{i}p_{st}({\bf m})=\frac{x}{1+x}=
\epsilon,
\end{equation}
and, consequently, the steady density of particles is $\rho_{st}=1-\epsilon$. As
$x$ increases, the equilibrium density of particles decreases.

 The quantity $x(\epsilon)$ can be related with a temperature parameter $T$ by
defining an energy $E({\bf m})$ for the system. A possible choice is
\begin{equation}
\label{2.10}
E({\bf m})=e_{0}\sum_{i=1}^{N}m_{i},
\end{equation}
where $e_{0}$ is a constant fixing the energy scale. If now the
distribution given by Eq.\ (\ref{2.7}) is identified with the equilibrium
canonical distribution, it is easily obtained that
\begin{equation}
\label{2.11}
x=e^{-\beta e_{0}},
\end{equation}
with $\beta=(k_{B}T)^{-1}$, $k_{B}$ being the Boltzmann constant. Thus the limit
$x \rightarrow \infty$ ($\epsilon\rightarrow 1$) is equivalent to $T\rightarrow
0^{-}$ and the limit $x \rightarrow 0$ ($\epsilon \rightarrow 0$) to $T
\rightarrow 0^{+}$.   A purely random distribution of particles and holes
corresponds formally to the equilibrium distribution for $\epsilon=1/2$ or
$T\rightarrow \infty$.

The transition rates given in Eq.\ (\ref{2.6}) define an irreducible Markov
process for $\epsilon>0$, except for the state with all the sites occupied by
particles that can not evolve. In the limit $N \rightarrow \infty$, the
probability of this state is negligible, and all the solutions of the master
equation relax to the steady distribution given by Eq.\ (\ref{2.7}) \cite{vk92}.
The situation is different in the no desorption limit $\epsilon=0$. The density
of particles cannot decrease, and all the states of the system having every hole
surrounded by two particles are absorbent; no evolution is possible from them.

Our one-dimensional lattice model can be regarded as a very simple picture of an
horizontal section of a real granular system, near the bottom of the container.
Consider first the freely evolving case. In a real granular medium, particles
cannot go up due to gravity. They can only go down, as long as there is enough
empty space in their surroundings. Therefore, the packing fraction grows until
the hard-core interaction prevents more movements of particles, and a
mechanically stable configuration is found. This situation is naively resembled
by the evolution of our model in the no desorption limit, $\epsilon=0$. Starting
from a given configuration of particles and holes,  the system evolves by means
of adsorption processes, occurring on those sites having at least one nearest
neighbor hole. This leads to a monotonic increase of the density until all the
holes become isolated, i.\ e.\ surrounded by two particles.

Next, suppose a granular system submitted to vertical vibration. During the
vibration, particles belonging to a low horizontal section can go up, making the
local packing fraction decrease. The hard-core repulsion is also fundamental in
the vibrated case, since particles always need enough free volume close to them
in order to move. In our model, these pulses are introduced by allowing
particles to be desorbed, but the dynamics is ``facilitated''. A particle can
only be adsorbed or desorbed if at least one of its nearest  neighbor sites is
empty. This is done to mimic the short-ranged dynamical constraints in the real
granular system. Of course, the relative magnitude $\epsilon$ of the desorption
rate and the pulse duration $t_0$ are the parameters characterizing the process.

Then, tapping processes have been modeled in our lattice system in the following
way. We started from a purely random configuration, i.\ e.\ the equilibrium
configuration for $\epsilon=1/2$. Then, the system was allowed to relax with
$\epsilon=0$ until reaching a steady metastable configuration, characterized by
all the holes being isolated, from which the system can not evolve any more.
This is a convenient initial state for the compaction experiment and corresponds
to the loosely packed conditions used in real laboratory experiments
\cite{KFLJyN95}. In this way the average initial density of particles in our
tapping process has been $\rho \simeq 0.7$.

Pulses are modeled by suddenly increasing the value of $\epsilon$ to a value
greater than zero. This is equivalent to increase the temperature of the system.
The duration of each pulse was $t_{0} \ll 1$. Between pulses the system relaxes
with no external excitation, i.\ e.\ with $\epsilon=0$. The waiting time between
consecutive pulses was much larger than the relaxation time needed for the
system to become trapped in a new metastable configuration. The density was
measured just before starting a new pulse. The whole process was designed to
mimic what is done in real experiments.

\section{Evolution with constant transition rates}
\label{sec3}

 In  this Section we will study the evolution equations that determine the
relaxation of the density. First, we will derive these equations for an
arbitrary value of the parameter $\epsilon$ characterizing the relative
probability of a desorption event. Secondly, we will analyze the free relaxation
without  desorption, i.\ e.\ in the limit $\epsilon=0$, and the effect of pulses
separately, taking into account that the final state for one of the processes
gives the initial condition for the other.

In the following we will restrict ourselves to homogeneous and isotropic states.
This requires to consider appropriate initial and boundary conditions, and it is
consistent with the qualitative picture depicted in the previous Section. It
will be assumed that the limit $N\rightarrow\infty$ has been taken. Let us
define probability distributions of groups of $r+1$ consecutive holes by
\begin{equation}
\label{3.1}
D_{r}(t)\equiv \langle m_{i}m_{i+1} \cdots m_{i+r}\rangle_{t} =
\sum_{\bf m} m_{i}m_{i+1} \cdots m_{i+r}\, p({\bf m},t).
\end{equation}
The homogeneity of the system implies that the above expression does not
depend on the starting site $i$ considered. Evolution equations for the
moments $D_{r}(t)$ are easily obtained from the master equation,
\begin{equation}
\label{3.2}
\frac{\partial}{\partial t} D_{0}(t)= \epsilon D_{0}(t)-D_{1}(t),
\end{equation}
\begin{equation}
\label{3.3}
\frac{\partial}{\partial t}D_{1}(t)= \epsilon\left[ D_{0}(t)
+C_{0,0}(t) \right]-D_{1}(t)-D_{2}(t),
\end{equation}
and
\begin{equation}
\label{3.4}
\frac{\partial}{\partial t} D_{r}(t)=- r D_{r}(t) -D_{r+1}(t)+\epsilon
C_{0,r-1}(t)
+\epsilon\left[ D_{r-1}(t) +\sum_{j=1}^{r-1} C_{j-1,r-j-1}(t) \right],
\end{equation}
for $r \geq 2$. Here we have introduced the probability distributions of two
groups of holes separated by a site
\begin{equation}
\label{3.5}
C_{r,s}(t)=\langle m_{i}m_{i+1}\cdots m_{i+r}m_{i+r+2}
\cdots m_{i+r+s+2}\rangle_{t}.
\end{equation}
Besides, from now on we use the dimensionless time scale defined by $t^{*}=\nu
t$, although the asterisk is omitted for the sake of simplicity. Again as a
consequence of homogeneity, the functions $C_{r,s}(t)$ do not depend on the site
$i$ taken as the origin to measure them. Moreover, isotropy implies the symmetry
property $C_{r,s}(t)=C_{s,r}(t)$. In Eq.\ (\ref{3.2}) it is seen that the time
evolution of the density of holes $D_{0}(t)$ involves the nearest neighbor pair
distribution of holes $D_{1}(t)$. When the equation (\ref{3.3}) for this latter
distribution is considered, the situation becomes more complex. Not only the
three consecutive hole distribution $D_{2}(t)$ shows up, but also the second
neighbor pair moment $C_{0,0}(t)$ appears.

On the other hand, the whole hierarchy of equations gets much simpler in the
limit $\epsilon\rightarrow 0$. As discussed in Sec.\ \ref{sec2} this is the no
desorption limit and corresponds to $T\rightarrow 0^{+}$ (very low
temperatures).  For $\epsilon=0$,  Eqs.\ (\ref{3.2})--(\ref{3.4}) reduce to
\begin{equation}
\label{3.6}
\frac{\partial}{\partial t} D_{r}^{(0)}(t)=-r D_{r}^{(0)}(t)- D_{r+1}^{(0)}(t),
\end{equation}
for  all $r$. Hereafter, the superindex $0$ indicates that a quantity is
evaluated in a system evolving with $\epsilon=0$. The hierarchy (\ref{3.6}) can
be easily solved by using, for instance, the generating function method
\cite{vk92}. We introduce a generating function
\begin{equation}
\label{3.7}
G^{(0)}(y,t)=\sum_{r=0}^{\infty}\frac{y^{r}}{r!}\,  D_{r}^{(0)}(t).
\end{equation}
From Eqs.\ (\ref{3.6}) it is obtained that $G^{(0)}(y,t)$ obeys the equation
\begin{equation}
\label{3.8}
\frac{\partial}{\partial t}G^{(0)}(y,t)+(y+1) \frac{\partial}{\partial y}
G^{(0)}(y,t)=0,
\end{equation}
whose solution is
\begin{equation}
\label{3.9}
G^{(0)}(y,t)=G_{0}\left[(y+1)e^{-t}-1 \right],
\end{equation}
where $G_{0}(y)=G^{(0)}(y,0)$  is the initial condition, that will be determined
by the final situation after a pulse.  This expression has been previously
obtained in a different context \cite{FyR96}. For large times $G^{(0)}(y,t)$
approaches the limit
\begin{equation}
\label{3.10}
G^{(0)}(y,\infty)=G_{0}(-1),
\end{equation}
and, consequently,
\begin{equation}
\label{3.11}
\lim_{t\rightarrow \infty} D^{(0)}_{0}(t)=G_{0}(-1), \quad
\lim_{t\rightarrow \infty} D^{(0)}_{r}(t)=0,
\end{equation}
$r \geq 1$. The last result reflects the property that for $\epsilon=0$ all the
holes are isolated in the long time limit, i.\ e.\ they are always between two
particles. Therefore, the probability of finding two consecutive sites with
$m_{i}=1$ is null. This is a general property that does not depend on the
initial conditions. Of course, the asymptotic value of $D_{0}^{(0)}=\langle
m_{i}\rangle^{(0)}$ is determined by the initial state of the system,  being
smaller than its  initial value. It must be noticed that the hierarchy
(\ref{3.6}) admits as a stationary solution any constant value for $D_{0}$ as
long as $D_{r,st}=0$ for $r\geq 1$.

An interesting particular case is when the system is at equilibrium with a given
value of $\epsilon >0$ before being suddenly  changed to $\epsilon=0$. In terms
of the temperature introduced in Sec.\ \ref{sec2} this is equivalent to a quench
of the system to $T=0^{+}$. The initial condition for this process is now (see
Eq.\ (\ref{2.7})),
\begin{equation}
\label{3.12}
D_{r}(0)=\epsilon^{r+1}.
\end{equation}
Then,
\begin{equation}
\label{3.13}
G_{0}(y)=\epsilon e^{\epsilon y},
\end{equation}
and Eq.\ (\ref{3.11}) yields
\begin{equation}
\label{3.14}
D_{0}^{(0)}(\infty)=\epsilon e^{-\epsilon}.
\end{equation}
For a purely random initial distribution ($\epsilon=1/2$) it is
$D_{0}^{(0)}(\infty) \simeq 0.3033$, i.\ e.\ less than one third of the sites
are empty in the final metastable state, characterized by a ``frozen''
configuration.

In the above discussion there was no need for considering the time evolution of
the distributions $C_{r,s}(t)$ defined in Eq.\ (\ref{3.5}). Nevertheless, it is
evident that in the limit $t \rightarrow \infty$, $C_{0,0}^{(0)}$ approaches a
constant value fixed by the initial conditions of the relaxation process, while
$C_{r,s}^{(0)}(t) \rightarrow 0$ for $r>0$ or $s>0$,  since the last ones
involve adjacent sites.

Next we analyze the evolution of the system with $\epsilon>0$ but for a time
interval $t_{0} \ll 1$.  This corresponds to the pulse preceding, and also
following, each of the free relaxations without desorption. Therefore, the
initial conditions we will be interested in correspond to a final state obtained
after a long time relaxation with $\epsilon=0$,
\begin{equation}
\label{3.15}
D_0(0)=m_{0}, \quad D_{r}(0)=0 \quad \text{for} \quad r\geq 1,
\end{equation}
\begin{equation}
\label{3.16}
C_{0,0}(0)=c_{0}, \quad C_{r,s}(0)=0 \quad \text{for} \quad r\geq 1 \quad
\text{or} \quad s\geq 1.
\end{equation}
For times $t \leq t_{0} \ll 1$ we approximate by means of a first-order Taylor
expansion using Eqs.\ (\ref{3.2})--(\ref{3.4}),
\begin{equation}
\label{3.17}
D_{0}(t) \simeq m_{0}+\epsilon m_{0}t,
\end{equation}
and similarly
\begin{equation}
\label{3.18}
D_{1}(t)\simeq \epsilon(m_{0}+c_{0})t,
\end{equation}
\begin{equation}
\label{3.19}
D_{2}(t) \simeq \epsilon c_{0}t,
\end{equation}
while distributions $D_{r}(t)$ with $r \geq 3$ are at least of order $t^{2}$.
Now, we can define the generating function corresponding to the pulse in a
similar way as it was done for $\epsilon=0$ in Eq.\ (\ref{3.7}). For this short
time limit it is
\begin{equation}
\label{3.21}
G(y,t)=\sum_{r=0}^{\infty} \frac{y^r}{r!} D_r (t) =m_{0}+t\epsilon
m_{0}+t\epsilon(m_{0}+c_{0})y+\frac{1}{2}t\epsilon c_{0}y^{2}+O(t^{2}).
\end{equation}

\section{Tapping processes}
\label{sec4}

 In this Section we will use the previous results to investigate the dependence
of the density on the number of taps. Let us consider the free relaxation with
$\epsilon=0$ after the $n+1$  pulse. The initial condition for this process will
be the final state reached during the pulse, i.\ e.\ using Eq.\ (\ref{3.21}),
\begin{eqnarray}
\label{4.1}
G_{n+1}^{(0)}(y,0) \equiv G_{0,n+1}(y)&=&m_{0,n}+t_{0}\epsilon
m_{0,n}+t_{0}\epsilon (m_{0,n}+c_{0,n})y    \nonumber \\
&&+\frac{1}{2}t_{0}\epsilon c_{0,n}y^{2}+O(t_{0}^{2}),
\end{eqnarray}
where $m_{0,n}$ and $c_{0,n}$ are the values of $\langle m_{i}\rangle$ and
$\langle m_{i}m_{i+2}\rangle$ at the end of the relaxation following the $n$-th
tap, respectively. The time evolution of the system during the relaxation is
described by Eq.\ (\ref{3.9}) and in the long time limit by Eq.\ (\ref{3.10}),
that particularized for the above initial condition yields
\begin{equation}
\label{4.2}
G_{n+1}^{(0)}(y,\infty) \simeq m_{0,n}-\epsilon c_{0,n} \frac{t_{0}}{2}
\end{equation}
and, therefore,
\begin{equation}
\label{4.3}
m_{0,n+1} \simeq m_{0,n}-\frac{1}{2} \epsilon t_{0} c_{0,n}.
\end{equation}
The above equation is expected to hold for small $t_{0}$ but arbitrary
``amplitude'' of the pulses $\epsilon$. Since $c_{0,n}$ is by definition
positive it follows that the density of holes decreases and the system
compacts monotonically as a function of the number of taps $n$. We stress
that the density is measured at the end of each free relaxation as it
is actually done in real experiments.

We have checked Eq.\  (\ref{4.3}) by comparing it with the results obtained from
Monte Carlo simulation of the Markov process defining the dynamics of the
system. An example is given in Fig. \ref{fig1} where we have plotted both
$2(\rho_{n+1}-\rho_{n})/\epsilon t_0$ and $c_{0,n}$ as functions of the number
of taps $n$. Here $\rho_{n}=1-m_{0,n}$ is the density of particles after the
$n$-th tap. In fact, since $m_{0,n+1}-m_{0,n}$  is a rapidly fluctuating
quantity, each of the points we have plotted corresponds to the average of those
functions over $10$ consecutive  taps. The data shown have been obtained in a
system of $10^{4}$ sites with $\epsilon=0.5$ and $t_{0}=0.02$, and have been
averaged over $10^3$ runs. It is seen that the prediction of the theory is
verified quite accurately. For the sake of clarity, we have restricted ourselves
to $5 \times 10^3$ taps, although the same behavior is observed until the system
comes near the steady state discussed in the next section.

Equation  (\ref{4.3}) indicates that the compaction process depends on the
product $\epsilon t_0$ and, in that sense, $\epsilon t_0$ plays in our model the
same role as $\Gamma$ in real experiments. The latter is defined as the ratio of
the peak acceleration of the tap to the gravitational acceleration
\cite{KFLJyN95}. Nevertheless, Eq.\ (\ref{4.3}) suggests a stronger prediction,
namely that the relevant time scale for the compaction process is
$\tau_{n}=\epsilon t_{0}n$. Of course, this will be true only if the dependence
of $c_{0,n}$ on $n$ also takes place through $\tau_{n}$, but it is easily seen
that it is really so. The initial condition for each pulse is a trapped
configuration that is metastable for $\epsilon=0$. That means that the
derivatives with respect to time of all moments are  proportional to $\epsilon$
at $t=0$, and in the limit of short duration pulses the change in any moment in
a pulse will be proportional to $\epsilon t_{0}$. This proportionality is
clearly kept by the free relaxation with $\epsilon=0$ that does not introduce
any new time scale in the problem. It is worth mentioning that the same results,
i.\ e.\ Eq.\ (\ref{4.3}), hold in the limit $\epsilon \rightarrow 0$, with
$\epsilon t_0 \ll 1$.

In Fig.\ \ref{fig2} the relaxation of the particle density is shown as
a function of the scaled time $\tau_{n}$ for different values of $\epsilon$
and $t_{0}$. In all cases $t_{0} \ll 1$ as required by the theory
we have developed. The system has $10^{4}$ sites. The initial state for all
the compaction experiments was the same, namely the fully random distribution.
As predicted, all points lie onto a single curve. Moreover, this scaled curve
is well described by the four-parameter  heuristic law
\begin{equation}
\label{4.4}
\rho_{n}=\rho_{\infty}-\frac{\delta \rho_{\infty}}{1+B \ln \left(1
+\frac{\tau_{n}}{\tau_{c}}\right)} \, ,
\end{equation}
with values of the parameters $\rho_{\infty}=1.10 $, $\delta \rho_{\infty}=
0.40$, $B=0.39$, and $\tau_{c}=3.37$. We do not observe any dependence of these
constants on the values of $\epsilon$ or  $t_{0}$, of course always in the limit
$t_{0} \ll 1$. The solid line in Fig.\ (\ref{fig2}) is the fit to Eq.\
(\ref{4.4}). We have tried to derive analytically a logarithmic law similar to
this empiric result, but we have not succeeded. It is not clear yet whether it
is just a convenient fitting expression with four parameters or it has a more
fundamental meaning, for instance associated to some peculiar dynamical events
which are dominant in the relaxation of the density. In this context, it is
important to realize that the law fails to describe the asymptotic behavior in
the limit of a large number of taps and the steady value of the density that is
eventually reached. In fact, the value of $\rho_{\infty}$ reported above is
clearly unphysical since it is larger than 1. We believe that this is a general
limitation of the law (\ref{4.4}) and it is not restricted to the present model.
We can substitute in Eq. (\ref{4.4})
\begin{equation}
\label{4.5}
\frac{\tau_{n}}{\tau_{c}}=\frac{n}{n_{c}},
\end{equation}
with $n_{c}=\tau_{c}/\epsilon t_{0}$. In this way the standard inverse
logarithmic law with time measured in number of taps is recovered
\cite{KFLJyN95}. But now we have an explicit dependence of $n_{c}$ on
$\epsilon t_{0}$. A similar result was found numerically in
Ref. \cite{CLHyN97} for a two-dimensional model with geometrical frustration.
Here the dependence appears as a consequence of the relevant scale
defining the time evolution of the system. This scale has been identified
by using analytical methods. The value $n_{c}$ can be understood as the
minimum number of taps needed to observe a significant compaction process. For
$n \ll n_{c}$ the density remains practically with its initial value.

\section{Steady states}
\label{sec5}
Another point we have investigated, prompted and stimulated by the results found
in previous works by different authors, is the possible existence of a long time
steady state density, determined by the tapping process (i.\ e.\ the amplitude and
duration of the pulses in the present model) but independent of the initial
conditions \cite{NKPJyN97}. Then, we have carried out a series of computer
experiments corresponding to the same values of $t_{0}$ and $\epsilon$ but to
different initial conditions and, in particular, to different values of the
initial density. The states we have chosen are equilibrium states, corresponding
to initial densities of particles $0.5$, $0.75$, $0.9$ and $0.995$. The results
are presented in Fig.\ \ref{fig3}. All densities tend in the long time limit to
the same steady value. The data in the figure have been obtained in a system
with $\epsilon t_{0}=0.03$, but the same qualitative behavior has been found in
all the studied cases. One important point to remark is that it is possible to
start from a density higher than the asymptotic one and then the density
decreases as the number of tapings increases. Once again, this behavior is
analogous to what is observed in real experiments. Average densities above the
so-called random close packing limit, which is much smaller than the crystalline
value, are not obtained even after extensive vibratory settling.

If we look for steady solutions of Eq.\ (\ref{4.3}), the only consequence we can
reach is that such state requires $c_{0}$ to be much smaller than $m_{0}$. A
more specific statement can be obtained by considering the next order in the
expansion of powers of $t_0$. Besides, we have simplified the calculations by
considering a pair approximation for all the correlation functions. More
specifically, we neglected all correlations involving more than two sites and
approximated in Eqs.\ (\ref{3.2}) and (\ref{3.3})
\begin{equation}
\label{5.1}
D_{2}(t)\equiv \langle m_{i}m_{i+1}m_{i+2}\rangle_{t}
\simeq \frac{D_{1}^{2}(t)}{D_{0}(t)},
\end{equation}
\begin{equation}
\label{5.2}
C_{0,0}(t) \equiv \langle m_{i}m_{i+2}\rangle_{t} =
\langle m_i m_{i+1}m_{i+2}\rangle_{t}+\langle m_i (1-m_{i+1}) m_{i+2} \rangle_t
 \simeq \frac{D_{1}^{2}(t)}{D_{0}(t)}+\frac{[D_{0}(t)-D_{1}(t)]^{2}}{1
 -D_{0}(t)}.
\end{equation}
The above approximations can be shown to be equivalent to the dynamical mean
field of clusters introduced by Dickman \cite{Di86}. When Eqs.\ (\ref{5.1}) and
(\ref{5.2}) are substituted into Eqs.\  (\ref{3.2}) and (\ref{3.3}), the latter
become a close pair of nonlinear first order differential equations,  namely
\begin{equation}
\label{5.2a}
\frac{dD_0(t)}{dt}= -D_1(t)+\epsilon D_0(t) ,
\end{equation}
\begin{equation}
\label{5.2b}
\frac{dD_1(t)}{dt}=-D_1(t)-\frac{D_1^2(t)}{D_0(t)}+\epsilon D_0(t)+
        \epsilon \left\{ \frac{D_{1}^{2}(t)}{D_{0}(t)}+
        \frac{[D_{0}(t)-D_{1}(t)]^{2}}{1-D_{0}(t)} \right\} .
\end{equation}
In the $n+1$ pulse we have to solve the above equations, for the time $t_0 \ll
1$ that the vibration lasts. Then, a perturbative solution in powers of $t_{0}$,
with the initial conditions given by
\begin{equation}
\label{5.3}
D_{0}(t=0)=m_{0,n}, \quad D_{1}(t=0)=0,
\end{equation}
is easily obtained,
\begin{equation}
\label{5.4}
D_{0}(t_{0})=m_{0,n}+\epsilon t_{0}m_{0,n}+\frac{t_{0}^{2}}{2} \left(
-\epsilon m_{0,n}+\epsilon^{2} m_{0,n}-\frac{\epsilon m_{0,n}^{2}}{1-m_{0,n}}
\right)+ O(t_{0}^{3}),
\end{equation}
\begin{equation}
\label{5.5}
D_{1}(t_0)=t_{0} \left( \epsilon m_{0,n} + \frac{\epsilon
m_{0,n}^{2}}{1-m_{0,n}}
\right)+\frac{t_{0}^{2}}{2} \left[-\epsilon m_{0,n}+\epsilon^{2}m_{0,n}
-\frac{\epsilon m_{0,n}^{2}}{1-m_{0,n}}
-\frac{\epsilon^{2} m_{0,n}^{3}}{(1-m_{0,n})^{2}} \right]+O(t_{0}^{3}).
\end{equation}
 Afterwards, the system evolves freely with $\epsilon=0$, and we measure the
density of holes $m_{0,n+1}$ at the end of this relaxation. We have to solve
Eqs. (\ref{5.2a})-(\ref{5.2b})  with $\epsilon=0$. Writing them as a closed
second order equation for the density of holes $D_0^{(0)}$, it is found that
\begin{equation}
\label{eq5.5a}
\frac{d}{dt} \left[ \frac{d \ln D_0^{(0)}(t)}{dt} + \ln D_0^{(0)}(t) \right]
= 0 ,
\end{equation}
and then
\begin{equation}
\label{eq5.5b}
\frac{d \ln D_0^{(0)}(t)}{dt} + \ln D_0^{(0)}(t)=
        -\frac{D_1(t_0)}{D_0(t_0)}+\ln D_0(t_0),
\end{equation}
since the initial conditions for the relaxation process are the final state of
the pulse, given by Eqs. (\ref{5.4}) and (\ref{5.5}). The value of the density
at the end of the relaxation is the long time limit solution of the above
equation, i.\ e.\
\begin{equation}
\label{5.8}
m_{0,n+1}=D_0^{(0)}(\infty)=D_{0}(t_{0})e^{-\frac{D_{1}(t_{0})}{D_{0}(t_{0})}},
\end{equation}
and using Eqs.\ (\ref{5.4}) and (\ref{5.5}),
\begin{equation}
\label{5.9}
m_{0,n+1}=m_{0,n}-\epsilon t_{0} \frac{m_{0,n}^{2}}{1-m_{0,n}}+\frac{1}{2}
(\epsilon t_{0})^{2} m_{0,n}\frac{1+m_{0,n}^{2}}{(1-m_{0,n})^{2}}+ O(t_{0}^{3}
m_{0,n}).
\end{equation}
As long as $m_{0,n}$ is much larger than $\epsilon t_{0}$ the third term on the
right hand side is negligible as compared with the second one, and the density
of holes decreases monotonically, i.\ e.\ the system is compacting. To the order or
approximation considered in Eq.\ (\ref{5.9}) a steady value of the density will
be reached when
\begin{equation}
\label{5.10}
m_{0,n+1}-m_{0,n}= O(t_{0}^{3}m_{0,n}).
\end{equation}
To get an expression for this steady density that we will denote by
$m_{0}^{(s)}$, let assume that it is of the order of $t_{0}^{\beta}$,
$m_0^{(s)}=c t_0^\beta$. Therefore,
\begin{equation}
\label{5.10a}
m_{0,n+1}-m_{0,n}=-\epsilon t_0^{1+2\beta} c^2+
\frac{1}{2} \epsilon^2 t_0^{2+\beta} c+O(t_0^{3+\beta})+O(t_0^{1+3\beta})+
O(t_0^{2+2\beta}),
\end{equation}
and a simple dominant balance of the first and second terms on the right hand
side of Eq.\ (\ref{5.9}) yields $\beta=1$ and $c=\epsilon/2$, i.\ e.\
\begin{equation}
\label{5.11}
m_{0}^{(s)}=\frac{1}{2} \epsilon t_{0}.
\end{equation}
We have discarded a solution $m^{(s)}=0$ that is always a trivial fix point for
the evolution of the system, corresponding to all sites being occupied by
particles. Let us notice that the above expression for $m_{0}^{(s)}$ is a steady
solution of the evolution equations in the pair approximation up to and
including order $t_{0}^{3}$.

In Fig.\ \ref{fig3} we have indicated the prediction of Eq.\ (\ref{5.11}). An
excellent agreement is found with the steady value reached in the simulations. A
comparison for several values of $\epsilon t_0$ is given in Fig.\ \ref{fig4}.
Again the theory describes fairly well, both qualitatively and quantitatively,
the results of the numerical simulations. The discrepancies increases as the
value of $\epsilon t_0$ increases, as expected. This confirms that in the
compaction experiment we are considering the steady correlations are determined
mainly by the two nearest neighbor (pair) correlations. This is due to the fact
that in the free relaxation processes the dynamics of the system is governed by
the restriction that there can not be a hole next to another one, but they are
isolated, i.\ e.\ surrounded by particles.

\section{Final remarks}
In the framework of a simple one-dimensional lattice model with facilitated
dynamics, we have studied the nonequilibrium evolution of a system
submitted to a tapping process. Trying to mimic what is done in real
experiments, the evolution of the system was modelled by a series of
two alternating steps. In the first one, the system evolves perturbed by
an external energy source, while in the second one it freely relaxes
towards a metastable configuration. The model has been shown to share
many of the characteristic features of granular materials under tapping.
In particular, the evolution of the density can be accurately described
by means of an inverse logarithmic law with the tapping number.

In our model , compaction is due to the decrease of the number of holes as the
system is being tapped. A qualitative picture of this very slow relaxation
follows from the existence of ``entropic barriers'' in the system. As the number
of holes lowers, the number of states allowing the system to relax become very
small as compared with the total number of available states. Therefore, it is
very difficult for the system to find the way to these bottlenecks in
configuration space, being most of the time trapped exploring metastable
configurations with almost the same density. This picture supports a strong
relationship between structural glassy relaxation and compaction \cite{kad99}.
In both cases there is a fast increase of the relaxation time of the system,
becoming very large on the time scale of the experiment.

Due to the simplicity of the model we have been able to obtain some detailed
analytical results. The process is characterized by the product $\epsilon t_0$,
that identifies the relevant time scale for densification, at least in the limit
of very short taps. Over this scale the time evolution of the system is
described by a universal law, which is independent of the particular values of
the parameters defining the system. This prediction has been compared with the
numerical solutions obtained by Monte Carlo simulation and a very good agreement
has been found.

A main result of this paper is an analytical expression for the asymptotic
density obtained in a tapping process. Quite surprisingly,  in the limit
considered there is a very simple proportionality relation between this density
and both the duration of the taps and their strength, the latter being measured
by the relative probability of a desorption event during a tap. Also this
theoretical prediction has been confirmed by the numerical solution.

\acknowledgments
This research was partially supported by the Direcci\'{o}n General de
Investigaci\'{o}n Cient\'{\i}fica y T\'{e}cnica (Spain) through Grant No.
PB96--0534.

\begin{figure}
\caption{
Plot of both $c_{0,n}$ (solid line) and $2(\rho_{n+1}-\rho_n)/\epsilon t_0$
(diamonds) as functions of the number of taps $n$, for $\epsilon=0.5$ and
$t_0=0.02$.}
\label{fig1}
\end{figure}

\begin{figure}
\caption{
Time evolution of the density of particles. Time is measured in the reduced
scale defined in the text. In the three curves shown, the values of the
parameters are $\epsilon=0.5$ in all of them and $t_0=2 \times 10^{-3}$
(diamonds), $t_0=0.01$ (squares), and $t_0=0.02$ (pluses).}
\label{fig2}
\end{figure}

\begin{figure}
\caption{
Evolution of the density as a function of the number of taps, for four different
values of the initial densities. The parameters characterizing the tapping
process are $\epsilon=0.5$ and $t_0=0.06$ in all cases.}
\label{fig3}
\end{figure}

\begin{figure}
\caption{
Stationary value of the density of particles, $\rho^{(s)}=1-m_0^{(s)}$, as a
function of the parameter $\epsilon t_0$ characterizing the tapping process. The
diamonds are numerical results, while the solid line is the analytical
expression given by Eq.\ (\protect\ref{5.11}).}
\label{fig4}
\end{figure}

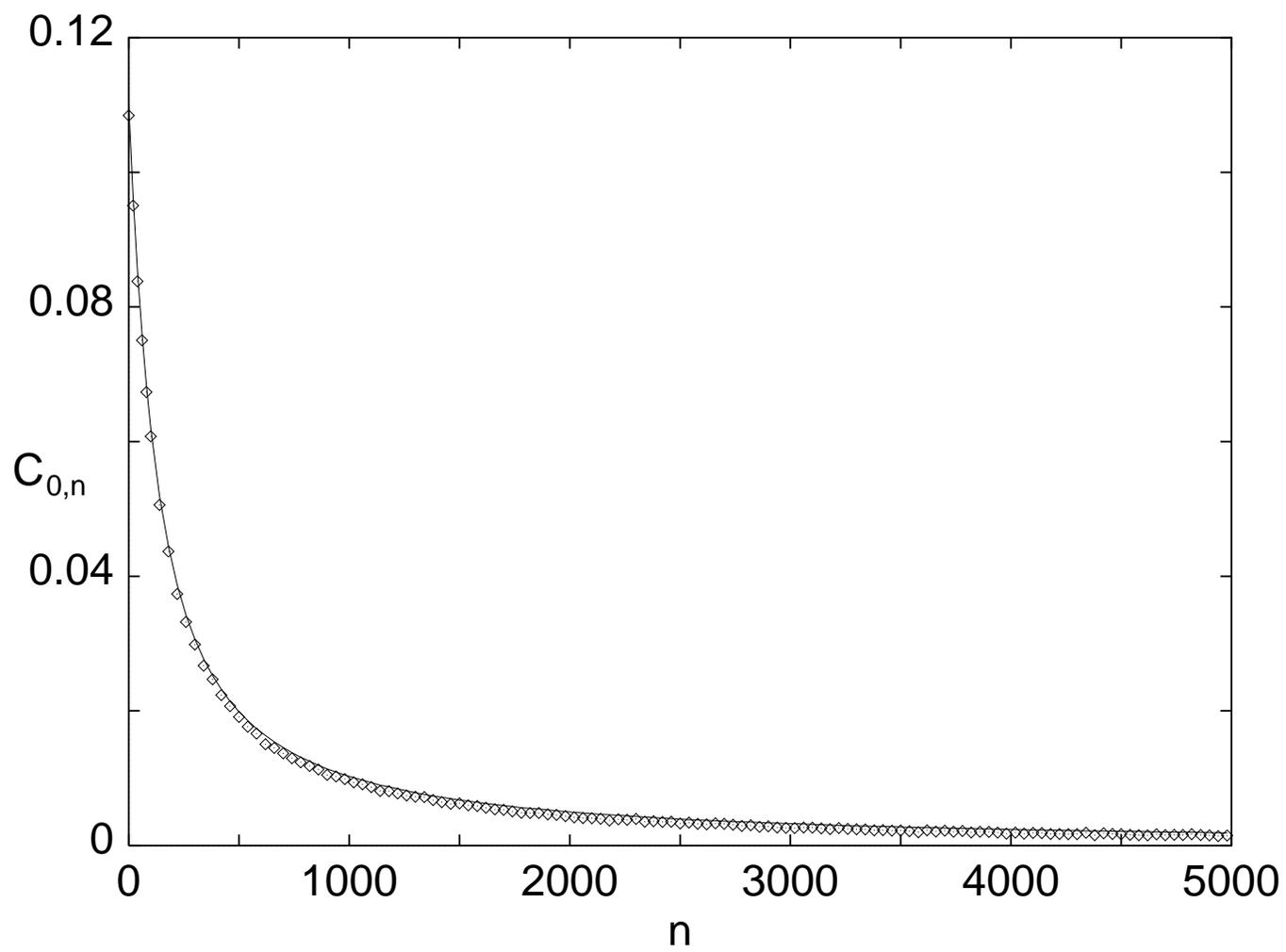
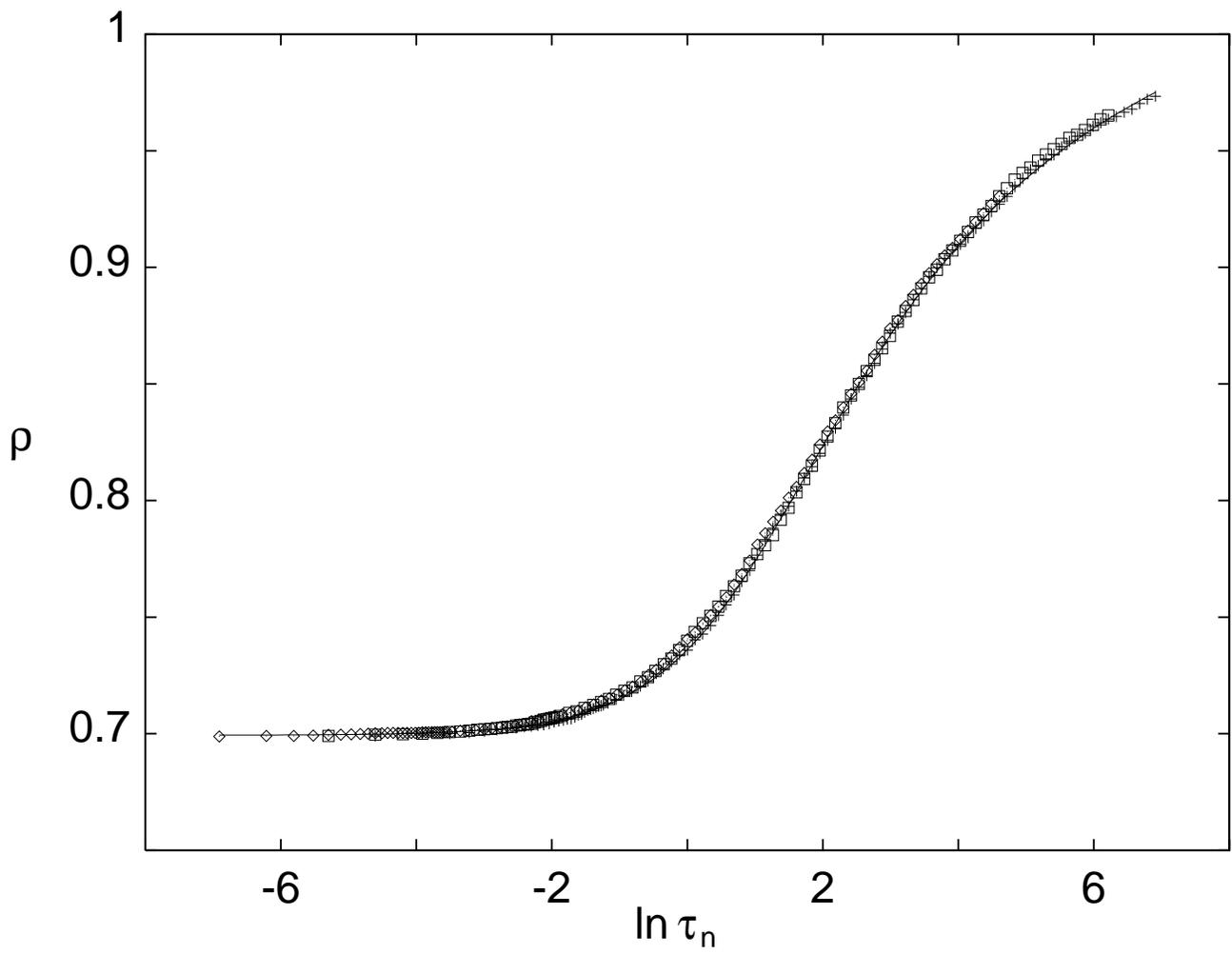
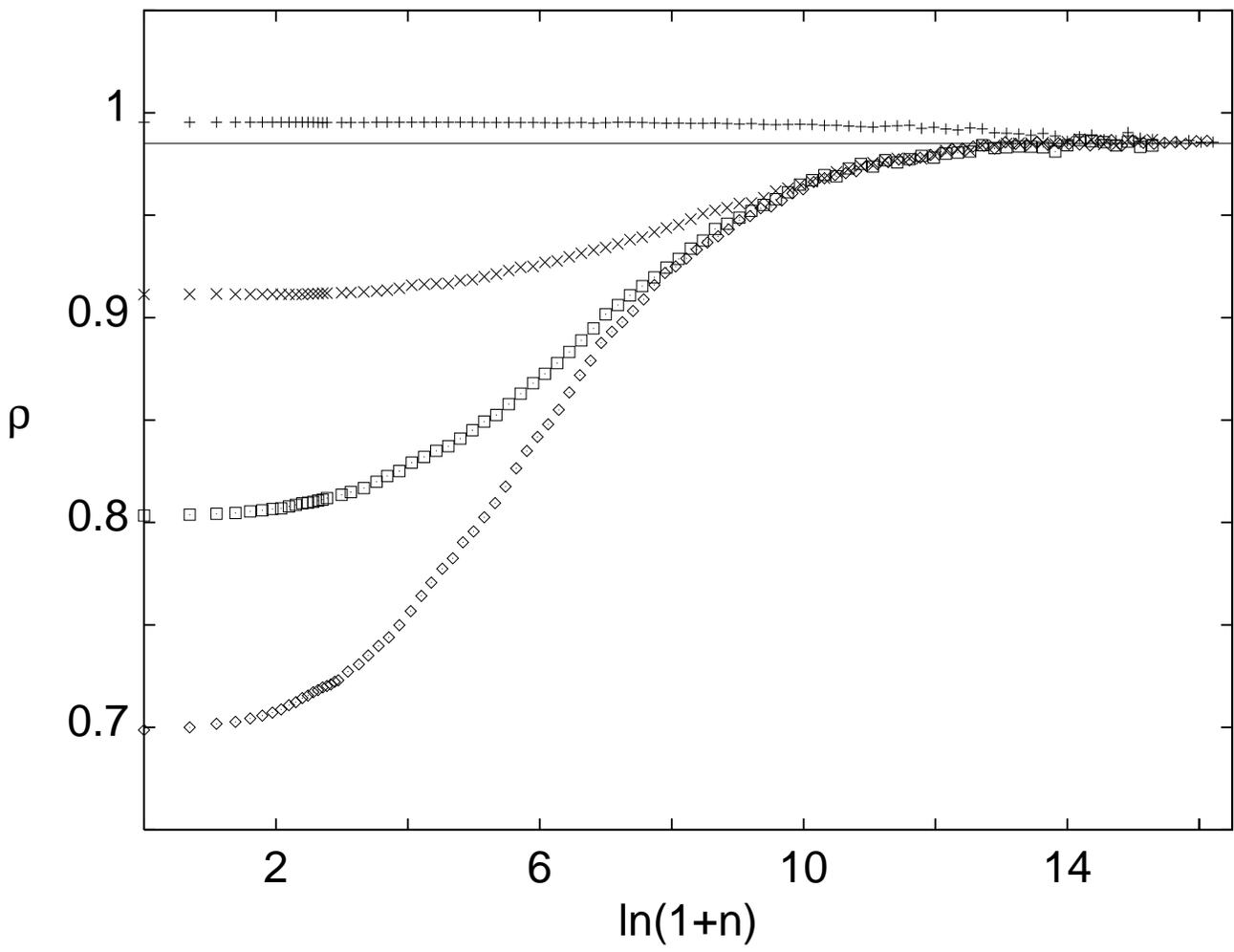
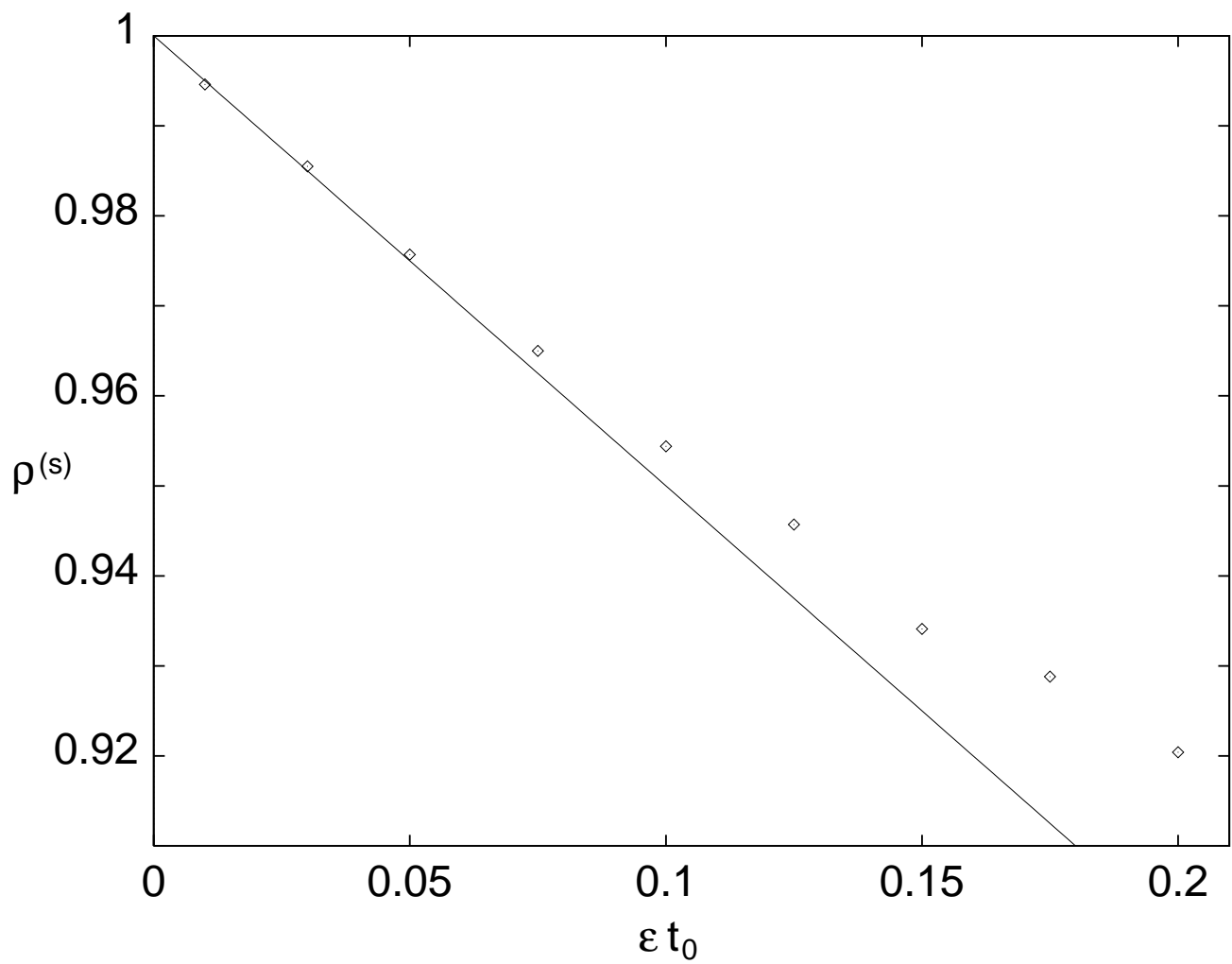
\end{document}